\documentclass{article}
\usepackage{amssymb}
\usepackage{amsfonts}
\usepackage{graphicx}

\input{tcilatex}

\begin{document}

\title{Reforming the international system of units: On our way to redefine
the base units solely from fundamental constants and beyond}
\author{ Ch.J. Bord\'{e}\thanks{%
Membre de l'Institut, Acad\'{e}mie des Sciences, President of the Conf\'{e}%
rence G\'{e}n\'{e}rale des Poids et Mesures (CGPM) in 1999, 2003, 2007 and
2011} \\
{\small \ Laboratoire de Physique des Lasers, UMR 7538 CNRS,\ Universit\'{e}
Paris Nord,}\\
{\small 99 avenue J.-B. Cl\'{e}ment, 93430 Villetaneuse, France}\\
{\small \ LNE-SYRTE, UMR 8630 CNRS, Observatoire de Paris}\\
{\small 61 avenue de l'Observatoire, 75 014 Paris, France}}
\date{}
\maketitle

\begin{abstract}
\begin{description}
\item Thanks to considerable progress in quantum technologies, the trend
today is to redefine all SI base units from fundamental constants and we
discuss strategies to achieve this goal. We first outline the present
situation of each of the seven base units and examine the choice of
fundamental constants which can reasonably be fixed. A critical issue is how
we should redefine the unit of mass in the context of modern relativistic
quantum theory. At the microscopic level the link of mass with proper time
as conjugate variables in the quantum phase $S/\hbar $ is well established.
This link strongly suggests that we should fix the value of Planck's
constant $h$, thus defining mass through a de Broglie-Compton frequency $%
mc^{2}/h$. This frequency can be accurately measured for atomic and
molecular species by atom interferometry. The main difficulty is then to
bridge the gap with the macroscopic scale for which phases are usually
scrambled by decoherence and where all mechanical quantities are built from
the classical action $S$ only without connexion to a quantum phase. Two ways
are now being explored to make this connexion: either the electric kilogram
which uses recent progress in quantum electrical metrology or atom
interferometry combined with the Avogadro number determination using a
silicon sphere. Consequences for a new definition of the unit will be
explored as the two methods hopefully converge towards an accurate value of
Planck's constant. Another important choice is the electric charge
connecting electrical and mechanical units: we could keep Planck's charge
and vacuum properties $\mu _{0}$ and $Z_{0}$, which is the case today or
shift to a fixed electric charge $e$ which seems to be the favourite choice
for to-morrow. We recall that temperature and time are linked through
Boltzmann's constant and there is a general agreement to fix that constant
after suitable measurements. Finally the unit of time is looking for a new
more universal and accurate definition based on Bohr frequencies
corresponding to higher and higher frequency clocks. A last challenge is to
produce a unified framework for fundamental metrology in which all base
quantities and relevant fundamental constants appear naturally and
consistently. We suggest a generalized 5D framework in which both
gravito-inertial and electromagnetic interactions have a natural geometrical
signification and in which all measurements can be reduced to phase
determinations by optical or matter-wave interferometry.
\end{description}
\end{abstract}

\section{Introduction: from the French Revolution to Max Planck.}

The metric system was born during the French Revolution with the idea of
settling a universal system of units, open to every people, in every time.
At that time, the dimensions of the Earth, the properties of water appeared
as a universal basis, but some time later James Clerk Maxwell judged them
less universal than the properties of the molecules themselves. The next
step was taken by George Johnstone-Stoney, then by Max Planck, showing that
a deeper aim was to found the system of units only on a set of fundamental
constants originating from theoretical physics. As a consequence, a long
divorce began between the practical requirements of instrumental metrology
and the dreams of theoretical physicists. Might they marry again? This has
now become possible thanks to a set of recent discoveries and new
technologies: laser measurements of length, Josephson effect, quantum Hall
effect, cold atoms, atom interferometry, optical clocks, optical frequency
measurements.... So a strong tendency to tie the base units to fundamental
constants is rising again, and the debate is open as to the relevance, the
opportunity and the formulation of new definitions.

Since the very beginning of this adventure the French Academy of Sciences
has had a leading role in the development of ideas and the settling of the
metric system. An Academy committee on Science and Metrology is still
working on this theme today. We outline in this paper some of the questions
under consideration in this committee and their underlying physical grounds.
Our purpose is to offer a logical analysis of the system of units and to
explore possible paths towards a consistent and unified system with an
original perspective. The path taken here builds on the fact that, thanks to
modern quantum technologies, any measurement can be reduced to a
dimensionless phase measurement thanks to optical or matter-wave
interferometry and we shall try to follow this simple guiding line. We shall
finally show how one could progress even further on the path of a synthetic
framework for fundamental metrology based upon pure geometry in five
dimensions. The reader who does not wish to enter into these mathematical
considerations can skip the last paragraph and Appendix 2. The conclusion
emphasizes the role of quantum mechanics and uncertainty relations in the
new metrology.
\section{Present status and evolution of the international system of units,
the SI.}

The SI (11th CGPM, 1960) comprises seven base units which are all more or
less concerned by the process of evolution mentioned above:

- the metre has already been given a new definition from the time unit and
the velocity of light in 1983 (see next section);

- the kilogram is still defined today by an artefact of iridium/platinum
alloy but as we shall see in detail it could find a new definition from
Planck's constant in a near future;

- the SI ampere is defined through a property of the vacuum, specifically
its magnetic permeability $\mu _{0}=4\pi .10^{-7}$ H/m (9th CGPM, 1948), but
the electrical units have \textit{de facto} already gained their
independence from the SI ampere, by adopting conventional values for
Josephson and von Klitzing constants and the natural temptation today is to
adopt the value of the electric charge $e$ in order to freeze the numerical
values of these constants;

- the kelvin is defined through the triple point of water, whereas fixing
Boltzmann's constant $k_{\text{B}}$ would be more satisfactory;

- the candela, unit of luminous intensity, is nothing else but a
physiological unit derived from an energy flux, hence redundant with the
other base units. Furthermore, it does not take into account the coherence
properties, spectral content and spatial mode content of the source. Hence
we shall not give any further consideration to this pseudo-base unit;

- the mole (added to the SI by the 14th CGPM in 1971) is defined from the
mass of the carbon atom by a dimensionless number, the Avogadro number. A
better determination of this number should give an alternative option in
which it would be fixed to redefine the mass unit from the mass of an atom
or of the electron. The tendency today is simply to retain this numerical
value as a conventional number of entities;

- the second, unit of time, was originally defined as the fraction 1/86 400
of the "mean solar day". The exact definition of the "mean solar day" was
left to astronomers. However, observations have shown that this definition
was not satisfactory owing to irregularities in the Earth rotation. To give
more accuracy to the definition, the 11th CGPM (1960) approved a definition
given by the International Astronomical Union based on the tropical year
1900. But experimental work had already shown that an atomic standard of
time, based on a transition between two energy levels of an atom or a
molecule, could be realized and reproduced much more accurately. Considering
that a more precise definition of the unit of time was essential for science
and technology, the 13th CGPM (1967/68) replaced the definition of the
second by the following (source BIPM) :

The second is the duration of 9 192 631 770 periods of the radiation
corresponding to the transition between the two hyperfine levels of the
ground state of the caesium 133 atom. It follows that the hyperfine
splitting in the ground state of the caesium 133 atom is exactly 9 192 631
770 hertz.$\ \ \ \ \ \ \ \ \ \ \ \ \ \ \ \ \ \ \ \ \ \ \ \ \ \ \ \ \ \ \ \ \
\ \ \ \ \ \ \ \ \ \ \ \ \ \ \ \ \ \ \ \ $

This definition refers to a caesium atom at rest at a temperature of 0 K.
This implies that the corresponding frequency should be corrected from
Doppler shifts and from the shifts coming from all sources of ambient
radiation (CCTF 1999).

The second should soon be better defined by an optical clock or even by a
much higher frequency clock (nuclear transition or matter-antimatter
annihilation process). Ideally, physicists would have dreamed of an atomic
hydrogen clock; it would have allowed to tie the time unit to the Rydberg
constant and possibly, some day, to the electron mass. But this choice could
be behind us today.

One should emphasize that the unit of time refers to proper time\textit{%
\footnote{%
We should carefully distinguish two different meanings of time: on the one
hand, time and position mix as coordinates and this refers to the concept of
time coordinate for an event in space-time, which is only one component of a
4-vector; on the other hand, time is the evolution parameter of a composite
system and this refers to the proper time of this system and it is a Lorentz
scalar (see below).}}. Proper time, as we shall discuss in detail, is
associated with the internal evolution of a massive object such as an atom
and is the measurable quantity from which the time coordinate is more or
less artificially constructed. The time unit definition should thus have
referred to the atom Bohr frequency and not to the radiation frequency.
Among other things no mention is made of the recoil shift.

So we are facing ill-assorted definitions piled up along the years, without
any global consistency. The direct connexion between the definition of a
base unit from a fundamental constant, the practical working out of it, and
a main scientific discovery is well illustrated by the case of the metre and
its new definition issued from the technological progress of laser sources.
This represents the archetype of the path to be followed for the other units.

\section{The example of the metre}

The metre is the best-known example of a base unit for which a new
definition was based upon a fundamental constant, the velocity of light in
vacuum $c$, thanks to progress in optics and especially laser physics during
the second half of the XXth century.

The coordinates of space and time are naturally connected by Lorentz
transformations within the conceptual frame of the theory of relativity, and
the velocity of light takes place as a factor of conversion in these
symmetry transformations. \textit{\ }The existence of a symmetry is a first
situation which allows to create an association between two units and hence
to reduce the number of independent units.

A second favourable condition is the existence of mature technologies to
implement this symmetry. Relativity uses clocks and rods to define time and
space coordinates. The rods of relativity are totally based on the
propagation properties of light waves, either in the form of light pulses or
of continuous beams whose frequency can now be locked to atomic clocks.%
\textit{\ }It was possible to redefine the length unit from the time unit,
because modern optics allowed not only the measurement of the speed of light
generated by superstable lasers with a relative uncertainty lower than the
best length measurements, but also because today the same techniques allow
the new definition of the metre to be realized in an easy and daily way.%
\textit{\ }

It is precisely optical interferometry and especially the work of Albert A.
Michelson (Nobel Prize winner in 1907) that allow us to go from the
nanometric length which is the wavelength linked to an atomic transition, to
a macroscopic length at the metre level. Michelson interferometers can
measure the tiniest length variations (10$^{-23}$) induced by gravitational
waves over distances ranging between hundreds of kilometres on earth and
millions of kilometres in spatial projects such as LISA. Any length
measurement can thus be reduced to a phase measurement i.e. to the
determination of an invariant number.

This evolution started in 1960 when the metre was redefined from the
radiation of the krypton lamp. The birth of lasers, in 1959, helped to carry
on steadfastly in that direction. Above all it was the discovery of
sub-Doppler spectroscopic methods, and particularly of saturated absorption
spectroscopy in 1969 \cite{BargerHall, Borde69} which turned lasers into
sources of stable and reproducible optical frequencies. The other revolution
was the technique of the MIM diodes introduced by Ali Javan that led to
measure the frequency of these light sources directly from the caesium
clock. From then on, the velocity of light could be measured with a
sufficiently small uncertainty, and so the CGPM in 1983 fixed its value
linking the metre to the second. All the above implies a procedure to put
the definition into practice (\textit{mise-en-pratique}), using wavelengths
of lasers locked to recommended atomic or molecular transitions.

Finally this redefinition was possible because there was a theoretical
background universally accepted to describe the propagation of light in real
interferometers.

To extend this approach let us investigate to what extent a similar
situation can be met for the other units and what fundamental constants are
available for each of them. A detailed discussion, partly reproduced in
Appendix 1, is given in references \cite{ChJB2004, RS}.

\section{The dimensioned and dimensionless fundamental constants and their
place in present physics:}

The fundamental constants we are referring to, come out of the major
theories of modern physics: relativity theory, quantum mechanics,
statistical mechanics, field theories, ....Consequently they rely on our
models and representations of the physical world.

What set of fundamental constants must we choose in the end ? They belong to
two very distinct categories. On the one hand, we have what can be called
conversion constants. Such constants are used to connect together quantities
originally believed to be of a different nature, but later understood to
refer to the same physical entity. A famous example is the equivalence
between heat and work which led to the mechanical equivalent of the calorie:
4.18 joules. The conversion constants have the dimension of the ratio
between the linked units. They can be given a fixed numerical value, and the
number of independent units is thus reduced. Several constants play this
role unequivocally: such was the case with the velocity of light, and it is
still the case with Planck and Boltzmann constants as discussed later. In
other cases we will have a choice to make between several constants of the
same nature: it will be the case of the electric charge for instance. On the
other hand, nature forces on us another sort of constants: the value of
non-dimensional ratios: such are, for example, the coupling constants linked
to the fundamental interactions. The best known are the fine structure
constant describing the coupling of matter with the electromagnetic field:%
\begin{equation}
\alpha =\frac{\mu _{0}ce^{2}}{4\pi \hbar }
\end{equation}%
and its gravitational analog

\begin{equation}
\alpha _{G}=Gm_{e}^{2}/\hbar c
\end{equation}%
involving the gravitation constant $G$\ and the electron mass $m_{e}$.

The value of these coupling constants cannot be discussed, and remains
independent of the system of units. It is a constraint to be taken into
account in our choices.

\section{The kilogram and the mole, determining Avogadro number with the
silicon sphere:}

Since 1889 (1rst CGPM) the mass unit has been the mass of the international
prototype, a platinum-iridium alloy cylinder baptised $\mathfrak{K}$ and
kept in a vault of the Pavillon de Breteuil with 6 copies. After the three
intercomparisons made in 1889, 1946/53 and 1989/92, there is now a general
agreement on the idea that the mass of the standard prototype, constant by
definition, has in fact drifted by several 10 or so micrograms (i.e. some 10$%
^{-8\text{ }}$in relative value). This situation in which the electrons and
other elementary particles of the universe have a mass value changing with
time, when the piece of metal in the vault in S\`{e}vres has not, is quite
embarassing. So, every effort must be done to modify the definition
(recommendation of the 21rst CGPM). It would be much more satisfactory and
justified to start from the mass of microscopic particles (electron or atom) 
\textit{a priori} quite reproducible, and then to climb up the macroscopic
scale. But if masses can be easily compared both at the macroscopic and at
the atomic scales, the connection between these two scales is quite
difficult. To make this connection we need to make an object with a known
number of atoms and whose mass could be compared to that of the standard
kilogram. This amounts to determining the Avogadro number $\mathfrak{N}_{%
\text{A}}$ which defines the mole. The mole is a quantity of microscopic
objects defined as a conventional number of identical objects. This number
(of course without dimension) has been arbitrarily chosen equal to the
number of supposedly isolated atoms, at rest and in their fundamental state,
contained in 0.012 kg of carbon 12. Consequently it is, up to a numerical
factor 0.012, the ratio of the mass of the standard prototype to the mass of
a carbon atom. Avogadro's constant $N_{\text{A}}$ generally refers to that
same number per mole, and it is expressed in mol$^{-1}$. This number and
this constant are just another way of expressing the mass of a carbon atom,
or its 12th part, which is the unified atomic mass unit \textit{m}$_{u}$.
\begin{figure}[h]
\begin{center}
\includegraphics[height=2.8988in]{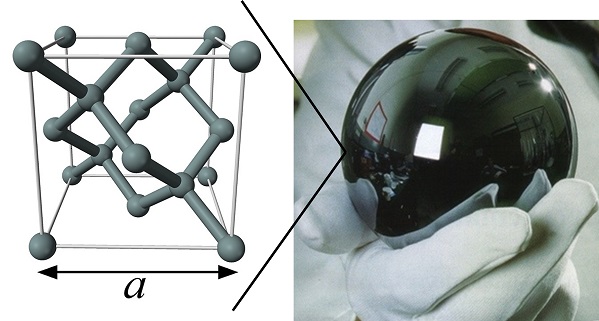}
\caption{Starting with a silicon
monocrystal purified by the floating zone method, several nearly perfect
spheres with masses $\sim 1$ kg were made (surface defects below some tens
of nanometres) then, thanks to mass spectrometry, X and optical
interferometries, the size $a$ of the cell $d_{220}$, the density $\protect%
\rho =V/m$ and the molar mass $M$ were determined. The cubic crystal cell
corresponding to eight atoms it was possible to obtain the Avogadro constant
from the formula $N_{A}=8M/(\protect\rho a^{3})$.} 
\end{center}
\end{figure}

An international program (the XRCD program for X-ray crystal density
program) has been developed to determine the Avogadro number from the
knowledge of a silicon sphere studied under "every angle": physical
characteristics of dimension, mass, volume, cell parameter, isotopic
composition, surface state etc...(see figure 1). The International Avogadro
Coordination project is refining the application of the XRCD method to
isotopically enriched $^{28}\func{Si}$ spheres with the goal of reaching a
1.5 10$^{-8}$ relative standard uncertainty \cite{Avo1}. The Avogadro
constant $N_{A}$ based on these measurements is presently 6.022 140 76(19) 10%
$^{23}$ mol$^{-1}$\cite{Avogadro}. This program has already faced and
overcome many difficulties, and one day it should eventually reach the goal
of fixing the Avogadro number with an accuracy that allows a redefinition of
the kilogram.

\section{Mass concept from proper time: Relativity and Quantum Mechanics.}

In fact, the notion of mass does not boil down to that of a quantity of
matter and, if a redefinition of the mass unit from the mass of a reference
elementary particle goes the right way, it does not reduce the number of
independent units. However, there is a possibility, as in the case of the
metre, to link the mass unit to the time unit. Indeed the theory of
relativity allows us to identify the mass $m$ of an object with its internal
energy, according to the well-known relation $E=mc^{2}$. What is more, Louis
de Broglie, in his famous Note in 1923 \cite{DeBroglie}, teaches us that
this energy can be linked to the proper time $\tau $ of the object to
produce the phase of an internal oscillation. The product $mc^{2}\tau $ of
these two quantities is an action, which must be related to an elementary
action, Planck's constant $h$, to give the phase without dimension of that
oscillation $mc^{2}\tau /h$ (see Appendix 2). In other words the quantity $%
mc^{2}/h$ is a frequency which we shall call de Broglie-Compton frequency
(dBC)\footnote{%
There is presently no physical clock at the de Broglie-Compton frequency
although it appears quite possible in the future through stimulated
absorption and emission of photon pairs in the creation/annihilation process
of electron-positron pairs.}. This frequency can be indirectly measured in
the case of microscopic particles such as atoms or molecules by modern
techniques of atomic interferometry in which de Broglie waves are precisely
made to interfere. The first experiments of this type were performed by
measuring the recoil frequency shift which occurs when laser light is
absorbed or emitted by molecules in saturation spectroscopy (Hall and Bord%
\'{e}, 1973 \cite{Hall}). They were followed by cold atom interferometry 
\cite{Chu} using Bord\'{e}-Ramsey interferometers \cite{ChJB89,Borde1,Borde6}%
. Today this measurement is done with a relative uncertainty less than 10$%
^{-8}$(Biraben et al. \cite{Biraben,Biraben2}). From that point, by simply
multiplying with Avogadro number $\mathfrak{N}_{A}$, we can have access to
the de Broglie-Compton frequency of the kilogram from that of the atomic
mass unit $m_{\mathfrak{u}}$, 
\begin{equation}
\nu _{dBC}=\frac{M_{\mathfrak{K}}c^{2}}{h}=1000\mathfrak{N}_{A}\left( \frac{%
m_{\mathfrak{u}}c^{2}}{h}\right)
\end{equation}%
and so link the mass unit to the time unit. Then the mass unit would be
defined by fixing that de Broglie-Compton frequency, which amounts to fix
Planck's constant. Such was the recommendation made by the working group of
the Acad\'{e}mie des Sciences to the CIPM in 2005. The definition of the
unit of mass would essentially look like:

\textquotedblleft The kilogram is the unit of mass, it is the mass of a body
whose de Broglie-Compton frequency is equal to $%
(299~792458)^{2}/(6.6260693.10^{-34})$ hertz exactly\textquotedblright .
This definition has the effect of fixing the value of the Planck constant, $%
h $, to be $6.6260693.10^{-34}$ joule second exactly.

This definition currently meets several criticisms: besides being an
unfamiliar concept involving too large a number, it is a quantum-mechanical
concept used in a range where its validity may be questioned because of
decoherence among other things\footnote{%
A well-defined phase assumes that the object should be in an eigenstate of
its internal Hamiltonian. In the case of a collection of atoms this could be
realized only with a large Bose-Einstein condensate.}. There is certainly no
physical clock at such a high frequency, which thus appears as fictitious.
Even at the single atom level the connection between the de Broglie-Compton
frequency, which is measured indirectly, and a real clock frequency has been
the subject of recent controversy. We shall see in the generalized 5D
approach that the overall action and hence the overall phase cancels along
the classical trajectory. Interference fringes result only from the phase
added by a coupling of modes with different wave vectors or frequencies. A
real atomic clock is generated at the Bohr frequency by a superposition of
two internal states $b$ and $a$ and it oscillates at the difference of the
two corresponding de Broglie-Compton frequencies on both sides of an
interferometer:

\begin{equation}
\nu _{\text{Bohr}}=\frac{m_{b}c^{2}}{h}-\frac{m_{a}c^{2}}{h}
\end{equation}%
The unit of mass may now be defined from this difference of the de Broglie
-Compton frequencies of both states which has a clear physical signification
and, if the chosen transition is the atomic transition which defines the
unit of time, we make an explicit link between both units:

"The kilogram is the unit of mass, it is the mass of $N$ massive particles
without mutual interactions with a mass equal to the mass difference between
the two internal states which define the unit of time" where $N$ is a fixed
numerical value of $c^{2}/h\nu _{\text{Bohr}}$ obtained by fixing the value
of the Planck constant, $h$, to be $6.6260693.10^{-34}$ joule second exactly.

We shall come back on this point since, as we will see, another way of
measuring the de Broglie-Compton frequency of the kilogram exists; it uses
the spectacular progress of quantum electric metrology that we are now going
to recall.

\section{Quantum electric metrology: Josephson and quantum Hall effects}

The electrical units underwent two quantum revolutions at the end of the
previous century: the Josephson effect which allows us to realize the volt,
and the quantum Hall effect which allows us to carry out the ohm.
\begin{figure}[h]
\begin{center}
\includegraphics[width=3.in]{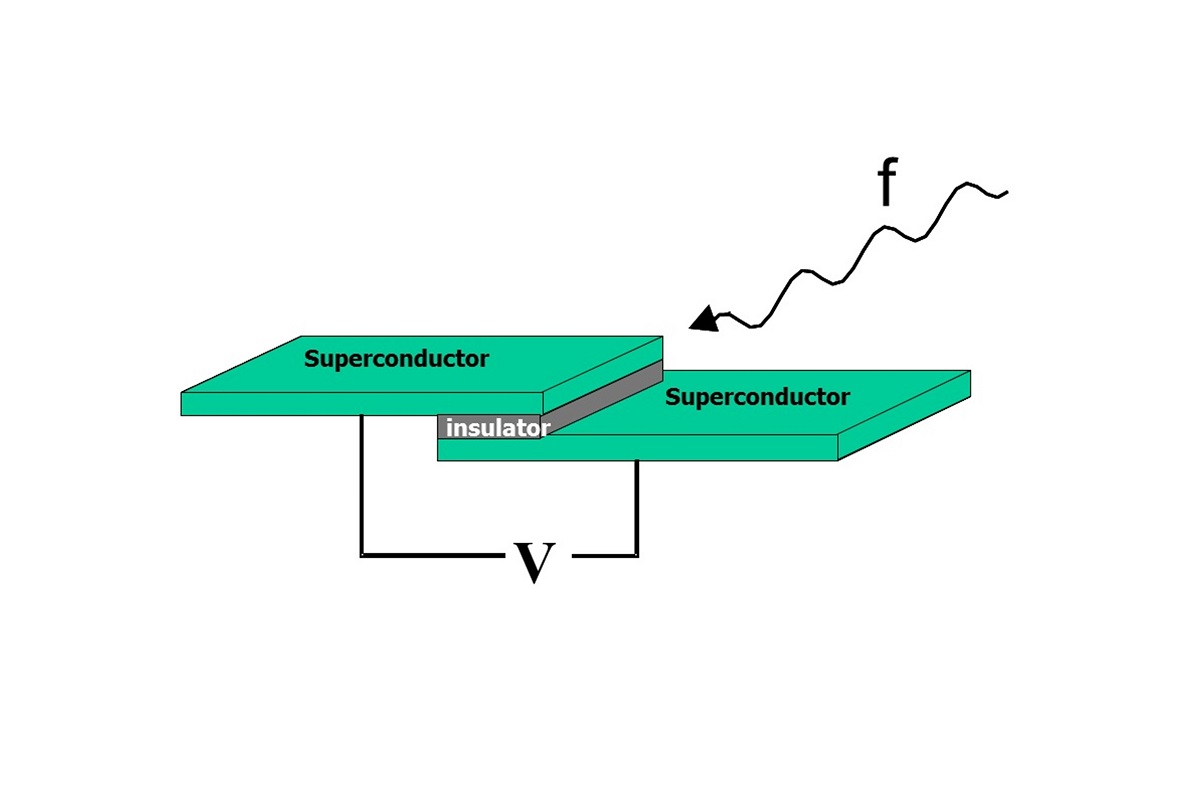}
\includegraphics[width=1.7in]{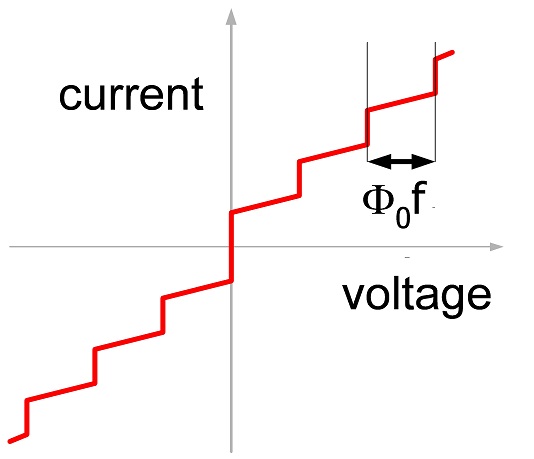}
\caption{The Josephson effect (Nobel prize 1973) uses the junction
comprising a very thin insulating layer sandwiched between two
supraconducting plates. When this junction is irradiated by an
electromagnetic wave of frequency $f$, its current-voltage characteristic
presents voltage plateaux connected to the frequency $f$ by a simple
proportionality relation in which $n$ is an integer characterizing each
plateau: $V=nK_{\text{J}}^{-1}f$.The Josephson constant $K_{\text{J}}$ is
given with an excellent approximation by $2e/h$. The charge $2e$ is that of
Cooper pairs of electrons which are able to tunnel across the junction. This
effect has a topological nature ($\protect\phi _{0}=h/2e$ is a quantum of
flux) hence its universal character, independent of the detailed realisation
of the junction and verified at the $10^{-10}$ accuracy level.}
\end{center}
\end{figure}

\begin{figure}[h]
\begin{center}
\includegraphics[width=1.9069in]{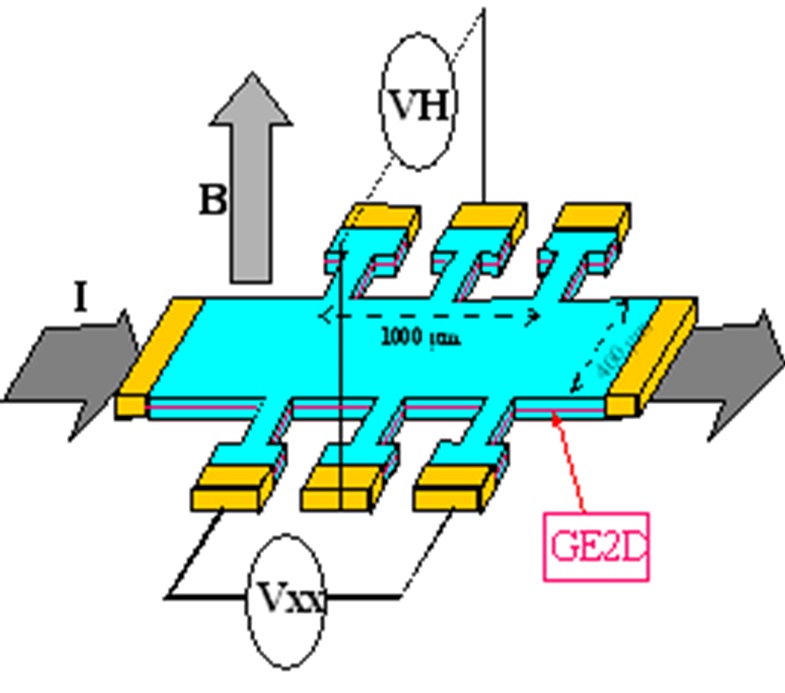}
\includegraphics[width=3.0632in]{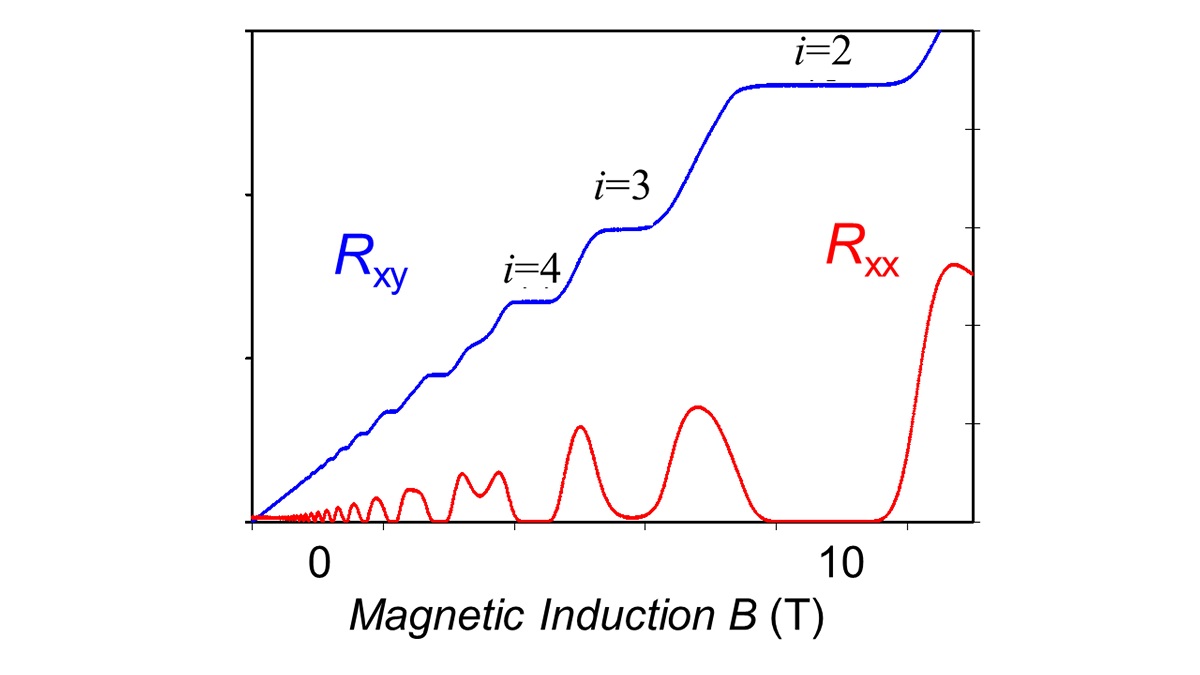}
\caption{Quantum Hall effect.
When a bidimensional gas of electrons in a semiconductor is submitted to a
strong magnetic field, the transverse resistance (Hall resistance) exhibits
steps quantized by the integer $i$ and von Klitzing (Nobel prize 1985)
resistance $R_{\text{K }}$whereas the longitudinal resistance vanishes: $R_{%
\text{H }}=R_{\text{K }}/i$. Here again the effect has a universal
topological nature and is protected by the chiral anomaly introduced by
Schwinger, which suggests that $R_{\text{K }}=h/e^{2}$ with an excellent
approximation.}
\end{center}
\end{figure}

Historically the ampere was the first example, before the metre, of a unit
defined from a fundamental constant, the magnetic permeability $\mu _{0}$ of
the vacuum (9th CGPM 1948). The combination of these two definitions fixes
all the propagation properties of electromagnetic waves in vacuum : velocity 
$c$ and impedance $Z_{0}=\mu _{0}c$. Let us remark that by fixing Planck's
constant, an electric charge would be also fixed, the Planck charge given by
: 
\begin{equation}
q_{\text{P}}=\sqrt{2h/Z_{0}}
\end{equation}

In practice the reproducibilities of Josephson and quantum Hall effects
(respectively $10^{-10}$ and $10^{-9}$ in relative value) reach such a level
that today electrical measurements use these effects without any other
connection to the definition of the ampere. Were Planck's constant fixed,
the electricians would be greatly tempted to fix the electron charge rather
than Planck's charge, having in the back of their mind to fix Josephson and
von Klitzing's constants. Unfortunately the simple theoretical expressions
that link these two constants to $e$ and $h$ have not yet been validated
with a high enough accuracy (only $2.10^{-7}$ for $K_{\text{J}}$ and $%
3.10^{-8}$ for $R_{\text{K }}$in relative value), even if their universality
could be demonstrated at a much higher level. Independently from the strong
theoretical arguments that lie under these formulas, it is necessary to make
sure that possible corrections are low enough for both effects to achieve a
reliable realization of $2e/h$ and $h/e^{2}$ (see figures 2 to 4).

In the case of the quantum Hall effect such a verification can be made
because the ratio of the vacuum impedance to $h/e^{2}$\ is just the double
of the fine structure constant $\alpha .$ The vacuum impedance can be
realized thanks to a calculable capacitor (Thomson-Lampard) and the
comparison of $Z_{0}/R_{\text{K }}$ with 2$\alpha $ value obtained by atom
interferometry presently sets an uncertainty level around $10^{-8}$ and will
certainly improve beyond $10^{-8}$. In the case of the Josephson effect the
limit comes from our insufficient knowledge of the proton gyromagnetic
ratio. Fortunately, two other verifications will be possible with the
metrologic triangle and the watt balance as we shall see below.

In fact, quantum electrical metrology is undergoing a third revolution with
the SET (Single Electron Tunnelling) permitting to count electrons one by
one. Then Ohm's law becomes an equality between frequencies: the electric
potential difference is turned to a Josephson frequency, the electric
current to a number of electrons by second, and the electric resistance
expressed in terms of von Klitzing resistance has no dimension (see figure
5).

The closure of the metrological triangle will be a real test of the quantum
realizations and of the theories that connect $K_{\text{J}}$ and $R_{\text{K 
}}$to the fundamental constants of physics. Presently it is done at some $%
10^{-7}$ level, but hopefully that limit will reach the $10^{-8}$ level in
the future.

Thus electrical metrology is in profound evolution. In the future it will
occupy a key position for the entire metrology, especially thanks to the
"electric" kilogram (see below). As for the electrical units, the question
can be raised if one should fix the positron charge $e$ or rather Planck's
charge $q_{\text{P}}$? The ratio of both charges being the square root of
the fine structure constant, the corresponding uncertainty will be
transferred to the non-fixed charge. Some arguments inspired by the recent
theories of strings and an easier statement of gauge invariance point to the
first choice. Caution towards the formulae giving $K_{\text{J}}$ and $R_{%
\text{K}}$ speaks for the second choice, which goes back to keep the vacuum
impedance fixed as it is now. This choice has in fact already been made by
the CIPM and the last recommendations of the CGPM are in favour of fixing
the electric charge $e$.
\begin{figure}[h]
\begin{center}
\includegraphics[width=3.1773in]{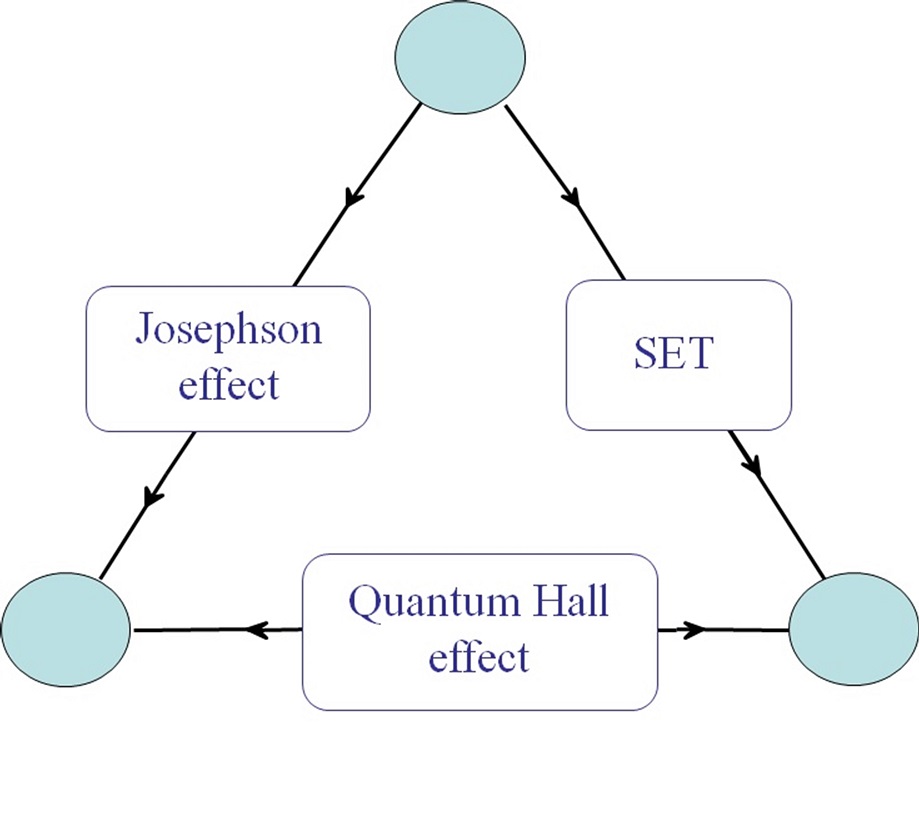}
\caption{The
metrological triangle is the quantum realisation of Ohm' law. If the three
effects are described by the canonical formulas with the same constants $e$
and $h$ in the three formulas: $U=\frac{h}{2e}f,I=ef^{\prime },\frac{R}{R_{%
\text{K}}}=\frac{e^{2}}{h}R$ one must check that $U=RI$ \ leads to an
equality between frequencies $f=2(R/R_{\text{K}})f^{\prime }$.}
\end{center}
\end{figure}

\section{The electric kilogram and the watt balance}

If the formulae giving $K_{\text{J}}$ and $R_{\text{K}}$ are considered to
be valid, Josephson and quantum Hall effects can be combined to produce an
electric power proportional to Planck's constant (figure 6): 
\begin{equation}
UI=\left( h/2e\right) ^{2}/\left( h/e^{2}\right) f_{1}f_{2}=\frac{h}{4}%
f_{1}f_{2}
\end{equation}%
where $f_{1}$and $f_{2}$ are the two Josephson frequencies involved in this
measurement.

This opened a new way of measuring the de Broglie-Compton frequency of the
kilogram, that of the "electric" kilogram. The "electric kilogram" was born
with the watt balance, suggested by Kibble in 1975 \cite{Kibble}, which in
one step (cryogenic version of the BIPM) or two steps (see figure 6),
carries out the direct comparison between a mechanical watt realized by
moving a mass in the gravitational field of the earth, and an electric watt
given by the combination of Josephson and quantum Hall effects. Such a
method demonstrated more than 20 years ago in the USA and in Great Britain
that it could reach a level of relative uncertainty consistent with that of
the present kilogram, i.e. some 10$^{-8}$. Two new realizations have been
assembled and are under study, one in Switzerland and a newer one in France.
Other programs will follow. Within a few years this effort is likely to
offer the opportunity to keep track of the evolution of the present kilogram
prototype, and later on to give a new definition of this kilogram by fixing
its dBC frequency. Clearly there is a competition between two projects to
define the mass unit: in the first project Planck's constant is fixed and
the watt balance allows to measure masses easily; in the second one
Avogadro's number is determined and fixed, and the mass unit is defined from
an elementary mass such as the electron mass. However, in that case, the
practical realization of a macroscopic mass still has to be done through the
realization of a macroscopic object whose number of microscopic entities is
known. The first point of view is the most attractive on the conceptual,
theoretical and even practical levels, even if the mass unit expression is
not easy to grasp for everyone. Anyway both ways towards Planck's constant
will need to be fully reconciled.

The CCM\footnote{%
Comit\'{e} consultatif pour la masse et les grandeurs apparent\'{e}es}(2013)
therefore \textit{recommends} that the following conditions be met before
the CIPM\footnote{%
Comit\'{e} international des poids et mesures} asks CODATA\footnote{%
Committee on Data for Science and Technology} to adjust the values of the
fundamental physical constants from which a fixed numerical value of the
Planck constant will be adopted,

1. at least three independent experiments, including work from watt balance
and XRCD experiments, yield consistent values of the Planck constant with
relative standard uncertainties not larger than 5 parts in $10^{8}$,

2. at least one of these results should have a relative standard uncertainty
not larger than 2 parts in $10^{8}$,

3. the BIPM prototypes, the BIPM ensemble of reference mass standards used
in the watt balance and XRCD experiments have been compared as directly as
possible with the international prototype of the kilogram
\begin{figure}[h]
\begin{center}
\includegraphics[width=4.4642in]{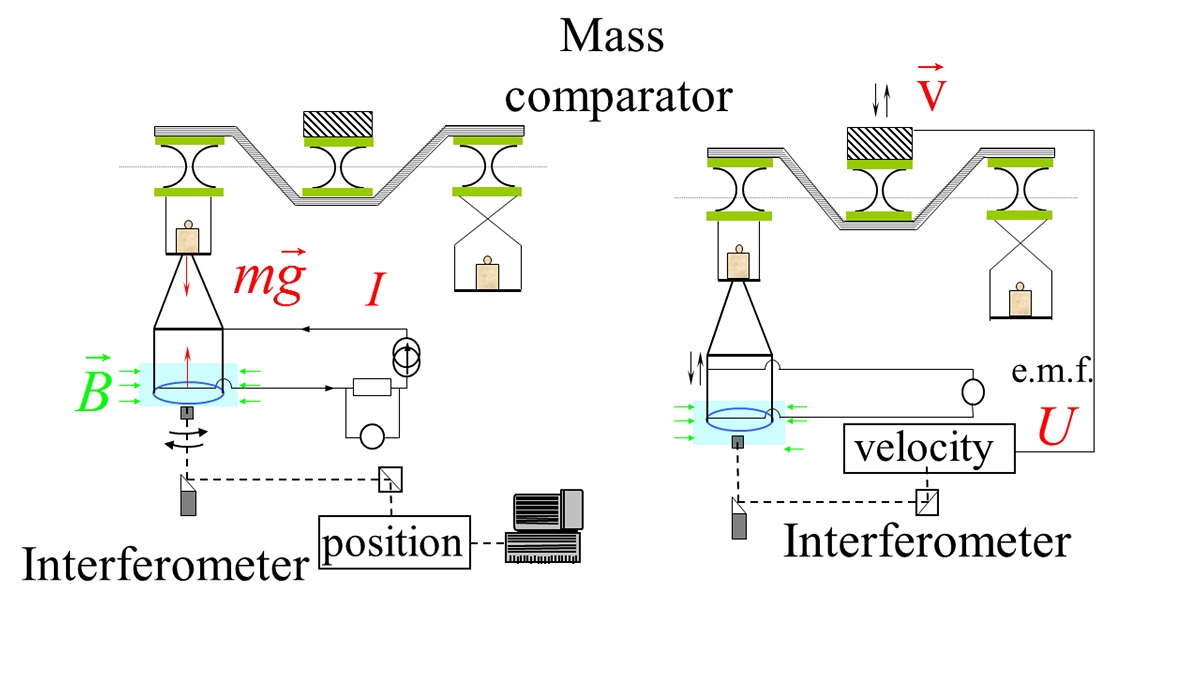}
\caption{In its classical version, the watt balance is
operated in two steps. In the first one, the weight of the kilogram in the
gravity field is balanced by the Laplace force exerted on a coil conducting
an electric current and placed in a magnetic field. The current $I$ is
measured by the combination of Josephson and quantum Hall effects. In the
second step the same coil is moved at constant speed v in the same magnetic
field and the induced emf $U$ is measured thanks to the Josephson effect.
Under these conditions, the properties of the coil and of the magnetic field
are common to both modes and the final formula expressing the equality
between electrical and mechanical powers $mgv=UI$ involves only times and
frequencies for the de Broglie-frequency of the kilogram $\protect\nu %
_{dBC}=M_{\mathfrak{K}}c^{2}/h=f_{1}f_{2}\protect\omega T^{2}/\left[ 4%
\protect\varphi (v/c)\right] $ where the Josephson frequencies of the two
steps are $f_{1}$ and $f_{2}$. The velocity $v$ is \ measured by optical
interferometry and the terrestrial gravity field.}
\end{center}
\end{figure}

There has been much progress in the determination of Planck's constant in
the recent past. A recent evaluation at NIST resulted in a value published
in 2014 with a standard uncertainty of $4.5$ parts in $10^{8}$. The last
result from NRC has a standard uncertainty of $1.8$ parts in $10^{8}$
sufficient to meet condition 2 of CCM. There are a number of other watt
balance experiments that will provide independent values.

We may imagine future ideal versions of the watt balance working as true
matter-wave interferometers analogous to superconducting ring gyros \cite%
{Zimmerman, Parker}. A superconducting coil directly connected to a
Josephson junction works as a Cooper pair interferometer which experiences a
phase shift from the change in gravitational potential. The power balance%
\begin{equation}
M_{\mathfrak{K}}gv=IU
\end{equation}%
may be written as

\begin{equation}
M_{\mathfrak{K}}c^{2}\left( \frac{gvT}{c^{2}}\right) =N_{2e}hf_{\text{J}}
\end{equation}%
in which $2eN_{2e}/T$ is the intensity of supercurrent, $v$ and $T$ are
respectively the velocity and the duration of the coil vertical motion. The
Josephson frequency $f_{\text{J}}$ thus appears as a measurement of the
gravitational shift of the dBC frequency of $N_{2e}$ Cooper pairs sharing
the mass $M_{\mathfrak{K}}$.

One can have a fascinating discussion on the question of whether quantum
mechanics applies or not when at the macroscopic scale of the kilogram and
on the real significance of the appearance of the Planck constant in the
watt balance formula. This debate has already started and must continue.
Whatever comes out, we are all already persuaded that the mass of a
macroscopic object is the sum of that of all its microscopic constituents
and of a weak approximately calculable interaction term. This hypothesis is
implicit in both possible new definitions of the unit of mass. The concept
of mass must be identical at all scales and mass is an additive quantity in
the non-relativistic limit. There is no doubt also that, at the atomic
scale, mass is directly associated with a frequency via the Planck constant.
This frequency can be measured for atoms and molecules even though it is
quite a large frequency. As mentioned earlier, measurements of $mc^{2}/h$
are presently performed with a relative uncertainty much better than $%
10^{-8} $. By additivity the link between a macroscopic mass and a frequency
is thus unavoidable. If one accepts to redefine the unit of mass from that
of a microscopic particle such as the electron, then the link with the unit
of time is \textit{ipso facto} established with a relative uncertainty much
better than $10^{-8}$.

Both units are \textit{de facto} linked by the Planck constant to better
than $10^{-8}$. It seems difficult to ignore this link and not to inscribe
it in the formulation of the system of units, especially since it leads to a
reduction of the number of independent units.

Another extremely important point is that mass is a relativistic invariant.
It should thus never be associated with the frequency of a photon field,
which transforms as the time component of a 4-vector in reference frame
changes. The de Broglie-Compton frequency is a proper frequency, Lorentz
scalar, equal by definition to $mc^{2}/h$.

Last, in the hypothesis of the mass unit being redefined by fixing Planck's
constant, the mole could be redefined separately from the kilogram by fixing
Avogadro's number. But should we not keep an exact molecular mass for carbon
12? If the mole is not any more directly connected to 12 grams of carbon,
its definition amounts to define an arbitrary number and this number cannot
be considered as a fundamental constant of nature. It is only if the mole
remains defined by 12 grams of carbon that it rests on a true physical
constant, the mass of the carbon atom. This constant has to be determined
experimentally if the unit of mass is defined by fixing the Planck constant.

\section{Boltzmann's constant and the temperature unit}

Statistical mechanics permits to go from probabilities to entropy thanks to
another dimensioned fundamental constant, Boltzmann's constant $k_{\text{B}}$%
. Presently the scale of temperature is arbitrarily fixed by the water
triple point, a natural phenomenon of course, but yet very far from
fundamental constants.

By analogy with the case of Planck's constant, it seems natural to propose
fixing Boltzmann's constant $k_{\text{B}}$. Indeed there is a deep analogy
between the two "$S$'s" of physics, which are action and entropy. They
provide respectively the phases and the amplitudes for the density operator.
The corresponding conjugate variables of energy are time and reciprocal
temperature with the two associated fundamental constants: the quantum of
action $h$ and the quantum of information $k_{\text{B}}$. Both participate
in statistical quantum mechanics through their ratio $k_{\text{B}}/h$. The
evolution parameter $\theta $ that comes in naturally \footnote{%
See for example the theory of linear absorption of light by gases and its
application to the determination of Boltzmann's constant (reference CRAS
2009).} in the combination of Liouville-von Neumann and Bloch equations for
the density operator $\rho $:%
\begin{equation}
i\hbar \frac{\partial \rho }{\partial \theta }=(H-<E>)\rho
\end{equation}%
is the complex time:%
\begin{equation}
\theta =t+i\hbar \beta /2=t+i\hbar /2k_{\text{B}}T
\end{equation}%
The link between atom interferometry and the Doppler broadening of line
shapes by the thermal motion of atoms is established in reference \cite%
{ChBLinAbs} which brings the connection between phase and temperature
measurements. The thermal motion of atoms is responsible for a loss of phase
coherence and the Doppler broadening may be seen as a limited visibility of
interference fringes.

An interesting analogy may be drawn for the two inaccessible limits that are
the velocity of light $c$ and the absolute zero temperature $T=0$. In both
cases the corresponding variable in $\theta $ becomes infinite. Internal
motion stops and both velocities $d\tau /dt$ (cf Langevin twins) and $u=%
\sqrt{2k_{\text{B}}T/m}\longrightarrow 0$ (The Doppler width and the black
body radiation shift vanish as the thermal decoherence time increases).

To measure Boltzmann's constant several methods, particularly acoustic
(propagation of sound in a gas), electrical (Johnson noise) and optical
(Doppler width measurements), are presently being studied \cite{CRAS1}. They
convey the hope of a low enough uncertainty (about 10$^{-6}$) to consider a
new definition of the kelvin from Boltzmann's constant later on. In
principle, such a redefinition does not face objections, and so it could be
done as soon as two different methods agree at the required accuracy. The
Boltzmann constant comes into play at the microscopic level through its
ratio to the Planck constant and at the macroscopic level through its
product by the Avogadro number. Any future redefinition of the kelvin should
take into account one of these associations, according to the future
definition of the unit of mass.

\section{What about the time unit? Towards a totally unified system?}

The measurement of time is the tip top of metrology. The accuracy of atomic
clocks has steadily increased by a factor 10 every ten years and this rate
has even accelerated recently with the advent of optical clocks \cite%
{Riehle, Abgrall, CRAS2}. Today their uncertainty reaches $10^{-18}$.
\begin{figure}[h]
\begin{center}
\includegraphics[width=5.8358in]{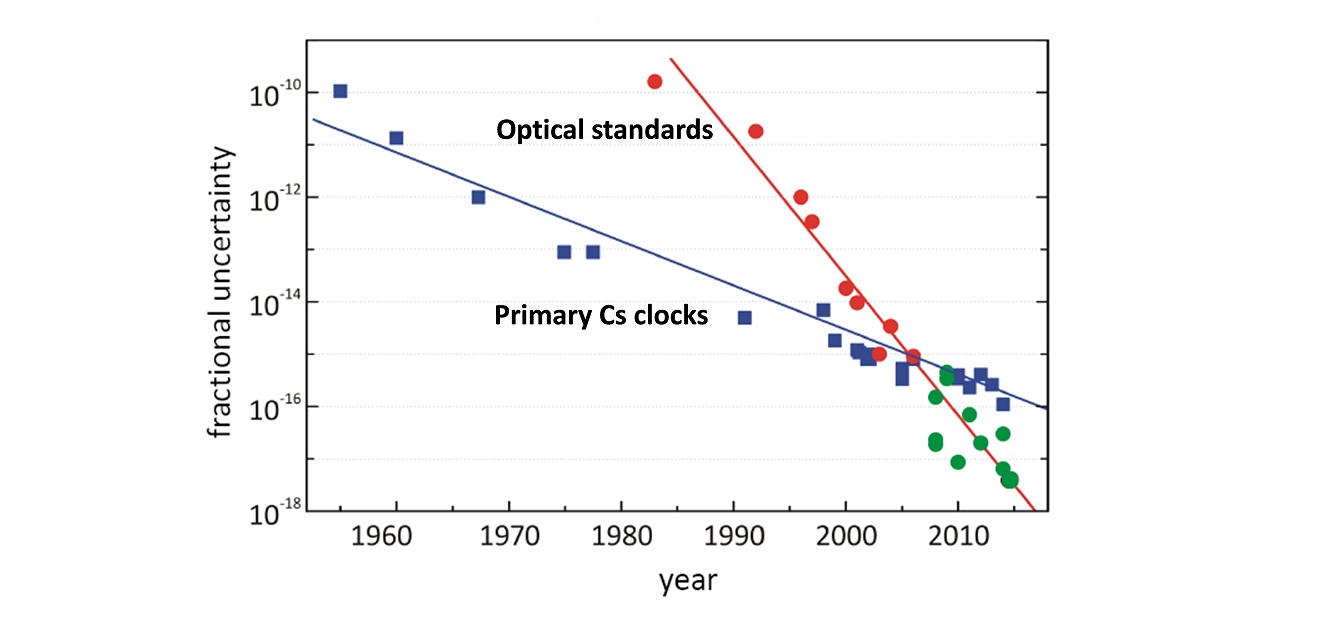}
\caption{Evolution of the frequency
accuracy of atomic clocks (from \protect\cite{Riehle})}
\end{center}
\end{figure}

 Thanks to this very high level of accuracy, time and
frequency measurements draw up the measurement of all other quantities. This
progress has its roots in the most recent atomic physics with cold atoms and
it finds everyday new applications, such as the global positioning satellite
system (GPS). The teams of SYRTE at the Paris Observatory and at the
Kastler-Brossel laboratory were pioneers in the use of cold atoms to create
clocks with atomic fountains. Among new revolutions we can quote the optical
clocks which, together with the frequency combs given by the femtosecond
lasers (J.L. Hall and T.W. Haensch Nobel prize 2005) will permit to count
better and faster, and there is every chance they will take the place of the
microwave clocks in the future. The competition runs high between neutral
atoms (in free flight or trapped in a light grating to benefit from the
Lamb-Dicke effect) and trapped ions. In the end, what part will the space
equipments play when it comes to compare clocks and to distribute time? In
the future the use of clocks on earth will inevitably be stopped at the
level 10$^{-17}$ by the lack of knowledge of the terrestrial gravitational
potential. Then an orbital reference clock will be needed. In the future who
will be the masters of time ?

The future possible redefinitions of the second are an open debate. Will the
second have, like the metre, a universal definition assorted with a way to
put it in practice, plus secondary realizations? This would raise, just as
in the case of the metre, the question of a possible variation of the
fundamental constants which would modify differently the different retained
transitions. The rubidium has better collisional properties than the caesium
and its hyperfine transition has been recommended by the CCTF (Consulting
Committee for Time and Frequency) as a secondary representation of the time
unit. On its side, hydrogen attracts many metrological physicists who would
like the definition of the time unit to be based on its transition 1s-2s.
That transition was the subject of spectacular intercomparisons (at 10$%
^{-14} $) with a cold Caesium fountain. A suitable combination of optical
frequencies could also be used to best isolate the Rydberg constant from
various corrections. The calculation of the hydrogen spectrum should be
carried as far as possible, at the same time keeping in mind the
considerable gap that will still separate theory from experience for a long
time. Last but not least, between the Rydberg constant and the electron mass 
$m_{e}$ we have the fine structure constant which is known only up to $%
0.7.10^{-9}$ so far, either by measuring the anomalous magnetic moment of
the electron or more recently by atom interferometry \cite{Biraben, Biraben2}%
. Obviously there is still a long way to formally tie the time unit to a
fundamental constant, but we must be aware of the implicit link existing
between the definition of the time unit and these fundamental constants. In
fact this situation is generic: owing to the permanent gap between theory
and a naturally very reproducible phenomenon we might never be able to
define units in terms of fundamental constants only. At some point we are
satisfied with formulas which describe the phenomenon until we discover
corrections that are too complex to be evaluated and we have then the choice
between having a simple definition from a fundamental theory or the use of a
complex but very reproducible experimental procedure. This is the situation
for the time unit now. In any case let us recall that the frequency provided
by an atomic clock should be corrected not only from the influence of all
external fields but also of Doppler and recoil shifts in order to yield a
true atom Bohr frequency and that such a Bohr frequency is the difference
between two de Broglie-Compton frequencies. Should Planck's constant be
fixed to define the unit of mass, the time unit would therefore always be
defined by the difference between two masses of an atomic species. It is our
choice to select either masses of elementary particles with the advantage of
simplicity or masses corresponding to internal states of very complex
objects far from fundamental physics but with the possible advantage of a
better reproducibility.

As a finishing touch to this quick survey of the base units and their
connexion with fundamental constants let us emphasize that a new metrology
in which quantum mechanics plays a more and more important part is building
up.

Presently, the whole scientific community is still urged to throw light on
the different choices aiming at the final decision at the CGPM of 2018.

\bigskip

\section{A generalized framework for fundamental metrology: 5D geometry
combining space-time and proper time}

Beyond the choice of relevant fundamental constants we must give coherence
to the new system of units. To obtain a consistent approach to this system
we must inscribe it in a unified physical framework for fundamental
metrology\ which contains a proper description of space-time, proper time
and mass and includes gravitation and electromagnetism as the main
interactions.\ This goal can be reached on purely geometrical\ grounds as we
show in Appendix 2.

\ This 5D scheme includes General Relativity with a 4D metric tensor $g^{\mu
\nu }$ and an electromagnetic 4-potential $\ A_{\mu }$ (with $\mu ,\nu
=0,1,2,3$) thanks to a metric tensor $G_{\hat{\mu}\hat{\nu}}$ for 5D such
that the generalized interval given by: 
\begin{equation}
d\sigma ^{2}=G_{\hat{\mu}\hat{\nu}}d\widehat{x}^{\hat{\mu}}d\widehat{x}^{%
\widehat{\nu }}\text{with }\hat{\mu},\hat{\nu}=0,1,2,3,4
\end{equation}%
is an invariant.

This metric tensor in five dimensions $G_{\hat{\mu}\hat{\nu}}$ is written as
in Kaluza's theory \cite{Kaluza} to include the electromagnetic gauge field
potential $A_{\mu }$: 
\begin{eqnarray}
G_{\hat{\mu}\hat{\nu}} &=&\left( 
\begin{array}{cc}
G_{\mu \nu } & G_{\mu 4} \\ 
G_{4\nu } & G_{44}%
\end{array}%
\right) =\left( 
\begin{array}{cc}
g_{\mu \nu }-\kappa ^{2}A_{\mu }A_{\nu } & -\kappa A_{\mu } \\ 
-\kappa A_{\nu } & -1%
\end{array}%
\right)  \nonumber \\
G^{\hat{\mu}\hat{\nu}} &=&\left( 
\begin{array}{cc}
G^{\mu \nu } & G^{\mu 4} \\ 
G^{4\nu } & G^{44}%
\end{array}%
\right) =\left( 
\begin{array}{cc}
g^{\mu \nu } & -\kappa A^{\mu } \\ 
-\kappa A^{\nu } & -1+\kappa ^{2}A^{\mu }A_{\mu }%
\end{array}%
\right)
\end{eqnarray}%
where $\kappa $\ is given by the gyromagnetic ratio of the object\footnote{%
In the case of the electron: $\kappa =e/m_{e}c$, which can also be written
as 
\[
\frac{1}{c}\sqrt{\frac{\alpha }{\alpha _{G}}}\sqrt{4\pi \varepsilon _{0}G} 
\]%
where we have introduced the dilaton field $\sqrt{\alpha /\alpha _{G}}$ to
make the connection with Kaluza's theory \cite{Kaluza}.}. It is such that
\bigskip \bigskip the equation :%
\begin{equation}
G^{\hat{\mu}\hat{\nu}}\widehat{p}_{\hat{\mu}}\widehat{p}_{\hat{\nu}}=0
\end{equation}%
with%
\begin{equation}
\widehat{p}_{\hat{\mu}}=(p_{\mu },-mc)
\end{equation}%
and \ $G_{44}=-1$ is equivalent to the usual \ equation in 4D for a massive
particle of mass $m$ and charge $q$:%
\begin{equation}
g^{\mu \nu }\left( p_{\mu }-qA_{\mu }\right) \left( p_{\nu }-qA_{\nu
}\right) =m^{2}c^{2}
\end{equation}%
These last equations give directly the Klein-Gordon equation in 5D for the
field $\varphi $:

\begin{equation}
\widehat{\square }\varphi =G^{\hat{\mu}\hat{\nu}}\widehat{\nabla }_{\hat{\mu}%
}\widehat{\nabla }_{\hat{\nu}}\varphi =0\text{ }
\end{equation}%
where the connection between mechanical quantities and quantum mechanical
operators is made as usual through Planck's constant. This equation is
analogous to the wave equation for massless particles in 4D.

The phase of this field is given by the 5D-superaction $\widehat{S}$ in
units of $\hbar $ and satisfies the Hamilton-Jacobi equation:%
\begin{equation}
G^{\hat{\mu}\hat{\nu}}\partial _{\hat{\mu}}\widehat{S}\partial _{\hat{\nu}}%
\widehat{S}=0\text{ }
\end{equation}

With this geometrical picture we have gathered all quantities concerned by
the main base units: space-time, proper time, mass, gravito-inertial and
electromagnetic fields in a phase without dimension. Any measurement can
then be reduced to a phase measurement through a suitable interferometry
experiment since all bases quantities enter the expression of a phase
through a comparison with reference quantities of the same nature. This
universal link is obtained by fixing Planck's constant. Mass and proper time
are entangled concepts which correspond to conjugate variables in classical
mechanics and to non-commuting operators in quantum mechanics in complete
analogy with momentum and position operators. The photon box of the
Einstein-Bohr controversy is a direct illustration of this quantum behaviour
and of the non-commuting character of proper time and mass operators:%
\begin{equation}
\left[ c\tau _{op},m_{op}c\right] =i\hbar
\end{equation}%
Their respective units thus require a joint definition in which the unit of
mass is defined from the mass difference of the two levels involved in the
definition of the unit of time. A compatible \textit{mise en pratique}
requires to associate a quantum clock with a macroscopic mass through a
phase measurement either by atom interferometry and atom counting or in the
watt balance. The Avogadro number is then obtained directly from the
measurement of the de Broglie-Compton frequency of the carbon atom in a
recoil experiment.

The proper time acquires a status in quantum mechanics and we may now
describe the quantum theory of atomic clocks in general relativity from
their internal properties since the phase of atom waves can be corrected
from general relativistic effects such as the gravitational red shift \cite%
{metrologia}.

Finally, temperature and time can be combined in a complex time variable in
the theory of clocks. This accounts for thermal decoherence through the
Doppler shift in atom interferometers. A generalized line shape for the
usual Doppler broadening can be derived accordingly \cite{ChBLinAbs}.

\section{Conclusion}

Most base quantities of metrology, length, time, mass, electrical
quantities, temperature are ultimately measured by optical or matter-wave
interferometers. Optics and quantum mechanics play a central role in the
description of these devices. As a consequence, future fundamental metrology
will deal essentially with phase measurements i.e. invariant numbers. One
should also emphasize the non-commuting character of quantities like mass
and proper time, which is a reason why Planck's constant has such a special
place in the system of units. Base quantities should be quantum observables.
Some appear as base quantities with their conjugate partner (e.g. mass and
proper time), others do not (e.g. position coordinate and momentum). The
quantum-mechanical link between conjugate quantities does not allow any more
to leave Planck's constant out of the system of units, which would be the
case if the mass unit continued to be defined by the mass of an object,
whether macroscopic ($\mathfrak{K}$) or microscopic (atom or electron). We
have seen that a natural choice was to couple the definitions for mass and
time units. Heisenberg's uncertainty relations will apply in the quantum
limit\footnote{%
One should also keep in mind the uncertainty relation between phase and
number of entities i.e. between action and quantity of matter.}. Measurement
theory becomes essential to explore the limits of the new quantum metrology.

A natural 5D theoretical framework for the redefinition of the SI is
completely provided by the connection between pure geometry, metric tensor
and metrology, that we have outlined. In this way, a clear separation has
been made between proper time (observable!) and time coordinate (not
observable!) as distinct quantities sharing the same unit. The role of the
electromagnetic field is to couple space-time and proper time coordinates
through the corresponding off-diagonal components of the metric tensor. The
5D action gathers all phenomena and constants of interest for a fully
relativistic quantum metrology in an invariant phase through Planck's
constant and this includes the dephasing arising from gravito-inertial
fields (e.g. the Sagnac effect or the effect of gravitational waves) as well
as those of electromagnetic origin (such as the Aharonov-Bohm or the
Aharonov-Casher effect).

Reduce the theory of measurements to the determination of quantum phases was
our primary objective and this paper is a first attempt to go in this
direction and to unify all aspects of modern quantum metrology. The
perspective that we have adopted, incorporates naturally all relevant
fundamental constants in a logical scheme with obvious constraints of
economy, aesthetics and rigour. The final aim is, of course, to adopt a
system of units free of arbitrary and artificial features, in harmony with
contemporary physics.

\section{Appendix 1: What framework for relativistic quantum metrology%
\textit{\protect\footnote{%
The following discussion is reproduced from references \cite{ChJB2004, RS}} }%
?}

This framework is naturally the one imposed by the two great physical
theories of the 20$^{th}$\ century: relativity and quantum mechanics. These
two major theories themselves have given birth to quantum field theory,
which incorporates all their essential aspects and adds those associated
with quantum statistics. The quantum theory of fields allows a unified
treatment of fundamental interactions, especially, of electroweak and strong
interactions within the standard model. General Relativity is a classical
theory, hence gravitation remains apart and is reintegrated into the quantum
world only in the recent theories of strings. We do not wish to go that far
and we will keep to quantum electrodynamics and to the classical gravitation
field. Such a framework is sufficient to build a modern metrology, taking
into account an emerging quantum metrology. Of course, quantum physics has
been operating for a long time at the atomic level, for example in atomic
clocks, but now it also fills the gap between this atomic world and the
macroscopic world, thanks to the phenomena of quantum interferences whether
concerning photons, electrons, Cooper pairs or more recently atoms in atom
interferometers \cite{ChJB89}.

The main point is to distinguish between a \textquotedblleft
kinematical\textquotedblright\ framework associated with fundamental
constants having a dimension, such as $c$, $\hbar $, $k_{\text{B}}$, and a
\textquotedblleft dynamical\textquotedblright\ framework where the
interactions are described by coupling constants without dimension. The
former framework relies on the Statistical Relativistic Quantum Mechanics of
free particles, and the latter on the quantum field theory of interactions.

Two possible goals can be pursued:

1 - redefine each unit in terms of a fundamental constant with the same
dimension e.g. mass in terms of the mass of an elementary particle

2 - reduce the number of independant units by fixing a fundamental constant
having the proper dimension for this reduction e.g. fixing $c$\ to connect
space and time units or $\hbar $\ to connect mass and time units.

The existence of fundamental constants with a dimension is often an
indication that we are referring to the same thing with two different names.
We recognize this identity as our understanding of the world gets deeper. We
should then apply an economy principle (Occam's razor) to our unit system to
take this into account and to display this connection.

When abandoning a unit for the sake of another, the first condition (C1) is
thus to recognize an equivalence between the quantities measured with these
units (e.g. equivalence between heat and mechanical energy and between mass
and energy), or a symmetry of nature that connects these quantities in an
operation of symmetry (for example a rotation transforming the space
coordinates into one another or of a Lorentz transformation mixing the space
and time coordinates).

A second condition (C2) is that a realistic and mature technology of
measurement is to be found. For example, notwithstanding the equivalence
between mass and energy, in practice the kilogram standard will not be
defined by an energy of annihilation, but on the other hand, thanks to the
watt balance, it can be tied to its Compton frequency $M_{\mathcal{K}%
}c^{2}/h $\ by measurements of time and frequency.

A third condition (C3) is connected to the confidence felt for the
understanding and the modelization of the phenomenon used to create the link
between quantities. For instance, the exact measurement of distances by
optical interferometry is never questioned because we believe that we know
everything, and in any case, that we know how to calculate everything
concerning the propagation of light. That is the reason why redefining the
metre ultimately took place without much problem. On the other hand,
measuring differences of potential by the Josephson effect or electrical
resistances by the quantum Hall effect, still needs support, because despite
a 10$^{-9}$\ confirmation of their reproducibility, and a good understanding
of the universal topological character behind these phenomena, some people
still feel uncertain as to whether all possible small\ parasitical effects
have been dealt with. For a physical phenomenon to be used to measure a
quantity properly, is directly related to our knowledge of the whole
underlying physics. In order to switch to a new definition, this
psychological barrier must be overcome, and we must have complete faith in
our total understanding of the essentials of the phenomenon. Therefore,
through a number of experiments as varied as possible, we must make sure
that the measurement results are consistent up to a certain level of
accuracy which will be that of the "mise en pratique" and we must convince
ourselves that no effect has been neglected at that level. If all of these
conditions are fulfilled, the measured constant that linked the units for
the two quantities will be fixed e.g. the mechanical equivalent of the
calorie or the speed of light.

\section{Appendix 2: The status of mass in classical relativistic mechanics:
from 4 to 5 dimensions}

In special relativity, the total energy $E$\ and the momentum components $%
p^{1},p^{2},p^{3}$ \ of a particle, transform as the contravariant
components of a four-vector

\begin{equation}
p^{\mu }=(p^{0},p^{1},p^{2},p^{3})=(E/c,\overrightarrow{p})
\end{equation}%
and the covariant components are given by : 
\begin{equation}
p_{\mu }=g_{\mu \nu }p^{\nu }
\end{equation}%
where $g_{\mu \nu }$ is the metric tensor. In Minkowski space of signature $%
(+,-,-,-)$:%
\begin{equation}
p_{\mu }=(p_{0},p_{1},p_{2},p_{3})=(E/c,-p^{1},-p^{2},-p^{3})
\end{equation}%
These components are conserved quantities when the considered system is
invariant under corresponding space-time translations. They will become the
generators of space-time translations in the quantum theory. For massive
particles of rest mass $m$, they are connected by the following
energy-momentum relation : 
\begin{equation}
E^{2}=p^{2}c^{2}+m^{2}c^{4}
\end{equation}%
or, in manifestly covariant form,%
\begin{equation}
p^{\mu }p_{\mu }-m^{2}c^{2}=0  \label{modulus}
\end{equation}%
This equation cannot be considered as a definition of mass since the origin
of mass is not in the external motion but rather in an internal motion. It
simply relates two relativistic invariants and gives a relativistic
expression for the total energy. Thus mass appears as an additional momentum
component $mc$ corresponding to internal degrees of freedom of the object
and which adds up quadratically with external components of the momentum to
yield the total energy squared (Pythagoras' theorem). In the reference frame
in which $p=0$ the squared mass term is responsible for the total energy and
mass can thus be seen as stored internal energy just like kinetic energy is
a form of external energy. Even when this internal energy is purely kinetic
e.g. in the case of a photon in a box, it appears as pure mass $m^{\ast }$
for the global system (i.e. the box). This new mass is the relativistic mass
of the stored particle:%
\begin{equation}
m^{\ast }c^{2}=\sqrt{p^{2}c^{2}+m^{2}c^{4}}
\end{equation}%
The concept of relativistic mass has been criticized in the past but it
becomes relevant for embedded systems. We may have a hierarchy of composed
objects (e.g. nuclei, atoms, molecules, atomic clock ...) and at each level
the mass $m^{\ast }$ of the larger object is given by the sum of energies $%
p^{0}$ of the inner particles. It transforms as $p^{0}$ with the internal
coordinates and is a scalar with respect to the upper level coordinates.

Mass is conserved when the system under consideration is invariant in a
proper time translation and will become the generator of such translations
in the quantum theory. In the case of atoms, the internal degrees of freedom
give rise to a mass which varies with the internal excitation. For example,
in the presence of an electromagnetic field inducing transitions between
internal energy levels, the mass of atoms becomes time-dependent (Rabi
oscillations). It is thus necessary to enlarge the usual framework of
dynamics to introduce this new dynamical variable as a fifth component of
the energy-momentum vector.

\bigskip Equation (\ref{modulus}) can be written with a five dimensional
notation :%
\begin{equation}
G^{\hat{\mu}\hat{\nu}}\widehat{p}_{\hat{\mu}}\widehat{p}_{\hat{\nu}}=0\text{
with }\hat{\mu},\hat{\nu}=0,1,2,3,4
\end{equation}%
where $\widehat{p}_{\hat{\mu}}=(p_{\mu },p_{4}=-mc)$ ;\ \ $G^{\mu \nu
}=g^{\mu \nu }$\ ;\ \ $G^{\hat{\mu}4}=G^{4\hat{\nu}}=0$ \ ;\ \ $G^{44}=\
G_{44}=-1$

This leads us to consider also the picture in the coordinate space and its
extension to five dimensions. As in the previous case, we have a four-vector
representing the space-time position of a particle:%
\[
x^{\mu }=(ct,x,y,z) 
\]%
and in view of the extension to general relativity:%
\begin{equation}
dx^{\mu }=(cdt,dx,dy,dz)=(dx^{0},dx^{1},dx^{2},dx^{3})
\end{equation}

The relativistic invariant is, in this case, the elementary interval $ds$,
also expressed with the proper time $\tau $\ of the particle:%
\begin{equation}
ds^{2}=dx^{\mu }dx_{\mu }=c^{2}dt^{2}-d\overrightarrow{x}^{2}=c^{2}d\tau ^{2}
\label{interval}
\end{equation}%
which is, as that was already the case for mass, equal to zero for light

\begin{equation}
ds^{2}=0
\end{equation}%
and this defines the usual light cone in space-time.

For massive particles proper time and interval are non-zero and equation (%
\ref{interval}) defines again an hyperboloid. As in the energy-momentum
picture we may enlarge our space-time with the additional dimension \ $%
s=c\tau $

\begin{equation}
d\widehat{x}^{\widehat{\mu }}=(cdt,dx,dy,dz,cd\tau
)=(dx^{0},dx^{1},dx^{2},dx^{3},dx^{4})
\end{equation}%
and introduce a generalized light cone for massive particles\footnote{%
In this picture, anti-particles have a negative mass and propagate backwards
on the fifth axis as first pointed out by Feynman. Still, their relativistic
mass $m^{\ast }$ is positive and hence they follow the same trajectories as
particles in gravitational fields as one can check from the equations of
motion \cite{Varenna}.}

\begin{equation}
d\sigma ^{2}=G_{^{\hat{\mu}\hat{\nu}}}d\widehat{x}^{\hat{\mu}}d\widehat{x}^{%
\widehat{\nu }}=c^{2}dt^{2}-d\overrightarrow{x}^{2}-c^{2}d\tau ^{2}=0
\end{equation}%
As pointed out in the case of mass, proper time is not defined by this
equation from other coordinates but is rather a true evolution parameter
representative of the internal evolution of the object. It coincides only
numerically with the time coordinate in the frame of the object through the
relation:

\begin{equation}
cd\tau =\sqrt{G_{00}}dx^{0}
\end{equation}

Finally, if we combine momenta and coordinates to form a mixed scalar
product, we obtain a new relativistic invariant which is the differential of
the action. In 4D:

\begin{equation}
dS=-p_{\mu }dx^{\mu }
\end{equation}%
and in 5D we shall therefore introduce the superaction:%
\begin{equation}
\widehat{S}=-\int \widehat{p}_{\hat{\mu}}d\widehat{x}^{\hat{\mu}}
\label{Superaction}
\end{equation}%
equivalent to%
\begin{equation}
\widehat{p}_{\hat{\mu}}=-\frac{\partial \widehat{S}}{\partial \widehat{x}^{%
\hat{\mu}}}\text{ \ with }\hat{\mu}=0,1,2,3,4
\end{equation}%
If this is substituted in%
\begin{equation}
G^{\hat{\mu}\hat{\nu}}\widehat{p}_{\hat{\mu}}\widehat{p}_{\hat{\mu}}=0\text{ 
}
\end{equation}%
we obtain the Hamilton-Jacobi equation in 5D%
\begin{equation}
G^{\hat{\mu}\hat{\nu}}\partial _{\hat{\mu}}\widehat{S}\partial _{\hat{\nu}}%
\widehat{S}=0\text{ }
\end{equation}%
which has the same form as the eikonal equation for light in 4D. It is
already this striking analogy which pushed Louis de Broglie to identify
action and the phase of a matter wave in the 4D case. We shall follow the
same track for a quantum approach in our 5D case.

What is the link between the three previous invariants given above? As in
optics, the direction of propagation of a particle is determined by the
momentum vector tangent to the trajectory. The 5D momentum can therefore be
written in the form:%
\begin{equation}
\widehat{p}^{\hat{\mu}}=d\widehat{x}^{\hat{\mu}}/d\lambda
\end{equation}%
where $\lambda $ is an affine parameter varying along the ray. This is
consistent with the invariance of these quantities for uniform motion.

In 4D the canonical 4-momentum is: $\ \ $%
\begin{equation}
p_{\mu }=mc\frac{g_{\mu \nu }dx^{\nu }}{\sqrt{g_{\mu \nu }dx^{\mu }dx^{\nu }}%
}=mcg_{\mu \nu }u^{\nu }
\end{equation}%
where $u^{\nu }=dx^{\nu }/d\tau $ is the normalized 4-velocity with $d\tau =%
\sqrt{g_{\mu \nu }dx^{\mu }dx^{\nu }}$ given by (\ref{interval}).

We observe that $d\lambda $ can always be written as the ratio of a time to
a mass:%
\begin{equation}
d\lambda =\frac{d\tau }{m}=\frac{dt}{m^{\ast }}=\frac{d\theta }{M}=...
\end{equation}%
where\ $\tau $ is the proper time of individual particles (e.g. atoms in a
clock or in a molecule), $t$ is the time coordinate of the composed object
(clock, interferometer or molecule) and $\theta $ its proper time; $m,$ $%
m^{\ast },M$ are respectively the mass, the relativistic mass of individual
particles and their contribution to the scalar mass of the device or
composed object.

From :%
\begin{equation}
d\sigma ^{2}=G_{^{\hat{\mu}\hat{\nu}}}d\widehat{x}^{\hat{\mu}}d\widehat{x}^{%
\widehat{\nu }}=0
\end{equation}%
we infer that in 5D%
\begin{equation}
d\widehat{S}=0
\end{equation}%
wheras in 4D%
\begin{equation}
dS=-p_{\mu }dx^{\mu }=-mc^{2}d\tau
\end{equation}

As a consequence the quantum mechanical phase also cancels along the
classical trajectory in 5D. The particle is naturally associated with the
position where all phases cancel to generate a constructive interference.

\ The previous 5D scheme can be extended to General Relativity with a 4D
metric tensor $g^{\mu \nu }$ and an electromagnetic 4-potential $\ A_{\mu }$ 
\cite{Varenna} with the metric tensor given in the main text.

\subsection{Acknowledgements}

This paper is adapted from an article in\ "La lettre de l'Acad\'{e}mie des
sciences" n$%
{{}^\circ}%
$20 (2007)\ which was kindly translated to English by Mrs Marie-Claire C\'{e}%
reuil. The author is also grateful to a number of colleagues for
constructive comments and especially Dr. Olivier Darrigol, Dr. Nadine de
Courtenay, Dr. Franck Pereira and Pr. Marc Himbert.

\end{document}